\documentclass[epj,nopacs]{svjour}
\usepackage{graphicx}
\usepackage{float}
\usepackage{amsmath,bm}
\usepackage{subfig}
\usepackage{setspace}
\usepackage{color}

\newcommand{\lra}[1]{\langle #1 \rangle }

\begin{document}
\title{Particle-laden two-dimensional elastic turbulence}
\author{Himani Garg\thanks{\email{himani.garg@etudiant.univ-lille1.fr}} \and Enrico Calzavarini \and Gilmar Mompean \and Stefano Berti
}                     


\institute{Universit\'{e} de Lille, Unit\'{e} de M\'{e}canique de Lille, EA , F-59000 Lille, France
}

\date{}

\abstract{ 
The aggregation properties of heavy inertial particles in the elastic turbulence regime of an Oldroyd-B fluid with 
periodic Kolmogorov mean flow are investigated by means of extensive numerical simulations in two dimensions.
Both the small and large scale features of the resulting inhomogeneous particle distribution are examined, focusing 
on their connection with the properties of the advecting viscoelastic flow. 
We find that particles preferentially accumulate on thin highly elastic propagating waves and that this effect is 
largest for intermediate values of particle inertia. We provide a quantitative characterization of this phenomenon 
that allows to relate it to the accumulation of particles in filamentary highly strained flow regions producing 
clusters of correlation dimension close to 1. 
At larger scales, particles are found to undergo turbophoretic-like segregation. Indeed, our results indicate 
a close relationship between the profiles of particle density and fluid velocity fluctuations. 
The large-scale inhomogeneity of the particle distribution is interpreted in the framework of a model derived in the limit 
of small, but finite, particle inertia. 
The qualitative characteristics of different observables are, to a good extent, independent of the flow elasticity. 
When increased, the latter is found, however, to slightly reduce the globally averaged degree of turbophoretic unmixing.  
}

\PACS{
     {}{}
     } 

\maketitle

\section{Introduction} \label{sec:1}
Viscoelastic fluids are known to be characterized by non-Newtonian behavior under appropriate conditions. In particular, 
dilute polymer solutions may display non-negligible elastic forces when the suspended polymeric chains occur to be sufficiently 
stretched by fluid velocity gradients. Remarkably, when the elasticity of the solution overcomes a critical value such forces can 
trigger instabilities that can eventually lead to irregular turbulent-like flow, even in the absence of fluid inertia, 
namely in the limit of vanishing Reynolds number. The latter dynamical regime is known as elastic turbulence \cite{GS00} 
and it has been experimentally observed in different flow configurations \cite{GS00,GS01,PMWA13,SACB17,sousa2018purely}.  
On the basis of its similarity with turbulent fluid motion, elastic turbulence has been proposed as an efficient system 
to enhance mixing in low Reynolds number flows \cite{GS01}. 
Moreover, it has been shown that it can increase heat transfer \cite{traore2015efficient,abed2016experimental}  
and promote emulsification \cite{poole2012emulsification}. 
Recently, it has also been argued that elastic turbulence flows play a significant role in the increased oil displacement obtained 
in industrial processes employing dilute polymer solutions to flood porous reservoir rocks \cite{MLHC16}.

Transport and mixing processes in fluids, however, often involve the presence of suspended finite-size impurities, 
like small and heavy solid particles. 
In view of mixing applications in elastic turbulence flows, it then seems necessary to 
accurately characterize how particle inertia affects the concentration of the transported species. 
Indeed, it is known that in turbulent flows the 
difference between the mass density of the impurities and 
that of the carrier fluid typically induces unmixing effects. Namely, it produces 
non-homogeneous particle distributions 
at small scales, as well as at large ones when turbulence spatial inhomogeneities are present (as, e.g., in a duct or in a boundary-layer flow). 
Although both types of inhomogeneities can be simultaneously present, they correspond to essentially different phenomena. 
While at small scales they give rise to complex clustered distributions due to the combined effect of small-scale turbulence 
and particle inertia~\cite{squires1991preferential}, at large scales they manifest in the accumulation of particles in regions of minimal 
turbulent intensity, whose locations are tightly related to the structure of the mean flow (turbophoresis) \cite{picano2009spatial,Sardina-2012,DCMB16}.

Inertial particle dynamics have been studied in turbulent flows of both Newtonian 
(see, e.g., \cite{squires1991preferential,bec2003fractal,calzavarini_kerscher_lohse_toschi_2008,toschi2009lagrangian}) 
and non-Newtonian (e.g., in \cite{DBM12,NSPB13}) fluids. 
The present work reports an investigation of heavy inertial particle transport at low Reynolds number, in a non-homogeneous flow 
of elastic turbulence in two dimensions. 
Despite the potential of the latter for mixing in microfluidics, 
the dynamics of particles in this regime are still quite unexplored.  
Our goal is to study the statistical features of particle aggregation at both small and large scales, and to relate them 
to the behavior of the main observables associated with the dynamics of the viscoelastic flow, as polymer elongation and velocity 
fluctuations, and to flow structures. 

The paper is organized as follows. The model used to describe the dynamics of heavy particles in a viscoelastic flow 
that can develop elastic turbulence states is introduced in Sec. \ref{sec:2}. In Sec. \ref{sec:3} we present the results 
of numerical simulations of this model. After briefly illustrating the main properties of the flow fields (Sec. \ref{sec:3.1}), 
we report on the properties of particle spatial distributions, separately focusing on preferential concentration 
effects (Sec. \ref{sec:3.2}), small-scale fractal clustering (Sec.~\ref{sec:3.3}) and large-scale inhomogeneities, 
i.e. turbophoresis (Sec. \ref{sec:3.4}). Conclusions are presented in Sec. \ref{sec:4}. 

 \section{Model of particle-laden viscoelastic flows} \label{sec:2} 
We consider the dynamics of a dilute polymer solution as described by the Oldroyd-B model~\cite{Bird1987}:
\begin{equation}
\partial_t \bm{u} + (\bm{u}\cdot\bm{\nabla}) \bm{u}  =  
- \frac{\bm{\nabla} p}{\rho_f} + \nu_s \Delta \bm{u} + {\frac{2\eta \nu_s}{\tau} \bm{\nabla} \cdot \bm{\sigma}} + \bm{f},
\label{eq:oldroyd_kf_u} 
\end{equation}
\begin{equation}
\partial_t \bm{\sigma} + (\bm{u} \cdot \bm{\nabla}) \bm{\sigma}  =  
(\bm{\nabla} \bm{u})^T \cdot \bm{\sigma} + \bm{\sigma} \cdot (\bm{\nabla} \bm{u}) - 2 \frac{\bm{\sigma}-\bm{1}}{\tau}.
\label{eq:oldroyd_kf_s}  
\end{equation}
In the above equations $\bm{u}$ is the incompressible velocity field, the symmetric positive definite matrix $\bm{\sigma}$ represents 
the normalized conformation tensor of polymer molecules and ${\bm 1}$ is the unit tensor corresponding to the equilibrium configuration 
of polymers attained in the absence of flow ($\bm{u}=0$). The trace $tr \left(\bm{\sigma}\right)$ gives the local polymer (square) elongation 
and $\tau$ is the largest polymer relaxation time. The fluid density is denoted ${\rho_f}$ and the total viscosity of the solution is 
$\nu = \nu_s(1 +\eta)$,  with $\nu_s$ the kinematic viscosity of the solvent and $\eta$ the zero-shear contribution of polymers (which is 
proportional to polymer concentration). The extra stress term ${\frac{2\eta \nu_s}{\tau} \bm{\nabla} \cdot \bm{\sigma}}$ accounts for 
elastic forces providing a feedback mechanism on the flow. 

In this study we are interested in having a fluid velocity characterized by a non-homogenous mean flow and turbulent fluctuations 
generated by elastic stresses only. For this reason we choose the two-dimensional (2D) periodic viscoelastic Kolmogorov flow. This  
has been previously shown~\cite{BBBCM08,BB10,PGVG17} to provide a simple and effective model 
able to reproduce the basic phenomenology of elastic turbulence.     
Using the Kolmogorov forcing $\bm{f} = (F\cos(y/L),0)$ in Eq. (\ref{eq:oldroyd_kf_u}), one has a 
laminar fixed point corresponding to the velocity field $\bm{u}^{(0)}=(U_0\cos(y/L),0)$ and the conformation tensor components 
$\sigma^{(0)}_{11}=1+\frac{\tau^2 U_0^2}{2L^2}\sin^2(y/L)$, 
$\sigma^{(0)}_{12}=\sigma^{(0)}_{21}=-\frac{\tau U_0}{2L}\sin(y/L)$, 
$\sigma^{(0)}_{22}=1$, with $F=  \nu U_0/L^2$ \cite{BCMPV05}. From these expressions, characteristic length and velocity scales 
$L$ and $U_0$, respectively, can be identified.
As previously documented, the laminar flow becomes unstable \cite{BCMPV05} for sufficiently high values of elasticity, 
even in the absence of fluid inertia, and eventually displays features typical of turbulent flows \cite{BBBCM08,BB10,PGVG17}. 
In the elastic turbulence regime, the mean velocity and conformation tensor fields keep similar trigonometric functional 
forms but with different amplitudes. 
Denoting $U$ the mean velocity amplitude in such states, we define the Reynolds number as $Re=UL/\nu$ and the Weissenberg number as $Wi=U\tau/L$, 
with their ratio giving the elasticity $El=Wi/Re$ of the flow.  

We assume that small spherical particles heavier than the fluid are laden in flows ruled by the dynamics described above. 
The suspension of impurities is considered to be dilute. The only force experienced by particles is Stokes drag and we then 
do not take into account the feedback effect of particles on the advecting fluid velocity and interactions among particles. 
Under these hypotheses the dynamics of each particle is described by the following equations of motion~\cite{maxey1983equation} 
for their position $\bm{x}$ and velocity $\bm{v}$: 
\begin{eqnarray}
\label{eq:part_x}
\dot{\bm{x}} & = & \bm{v}, \\
\label{eq:part_v}
\dot{\bm{v}} & = & -\frac{1}{\tau_p} \left[\bm{v}-\bm{u}(\bm{x},t)\right],
\end{eqnarray} 
where $\tau_p = \frac{2a_p^2\rho_p}{9\nu\rho_f}$ is the Stokes time, accounting for particle inertia, 
$a_p$ the particle radius, $\rho_p$ particle density (with $\rho_p \gg \rho_f$) and $\bm{u}(\bm{x},t)$ the advecting flow resulting 
from Eqs. (\ref{eq:oldroyd_kf_u}) and (\ref{eq:oldroyd_kf_s}). 
Particle inertia is typically parametrized by the Stokes number $St=\tau_p/\tau_f$ with $\tau_f$ a characteristic flow time scale. 
Here we define $\tau_f\equiv\tau_{\dot{\gamma}}$ in terms of the strain rate exerted by the flow, so that $St=\tau_p/\tau_{\dot{\gamma}}$, 
with $\tau_{\dot{\gamma}}=1/\overline{\dot{\gamma}}$ and $\overline{\dot{\gamma}}$ given by: 
\begin{equation}
 \overline{\dot{\gamma}} = \frac{1}{TL_0^2}\int_{0}^{T}dt\int_{0}^{L_0}dy\int_{0}^{L_0}dx \sqrt{2{[\bm{\nabla} \bm{u} +(\bm{\nabla} \bm{u})^T}]^2}, 
\label{eq:tau_gammadot}
\end{equation}
where $\overline{(...)}$ represents an average over spatial coordinates and time, $L_0$ being the domain size  
in each direction. 

\section{Analysis and Results }\label{sec:3}
To explore the dynamics of inertial particles in elastic turbulence we perform direct numerical simulations. 
Equations (\ref{eq:oldroyd_kf_u}) and (\ref{eq:oldroyd_kf_s}) are integrated using a pseudospectral method on a grid of side $L_0=2\pi$ 
with periodic boundary conditions at resolution $512^2$.  
Integration of viscoelastic models is limited by instabilities associated with the loss of positiveness of the conformation 
tensor \cite{sureshkumar1995effect}. These instabilities are particularly relevant at high $Wi$ values and limit 
the possibility to numerically investigate the elastic turbulence regime by direct implementation of the equations of motion. 
For this reason we adopt an algortithm based on a Cholesky decomposition of the conformation matrix 
ensuring symmetry and positive definiteness \cite{vaithianathan2003numerical}. This approach allows us to reach sufficiently 
high elasticity; however it imposes some limitations in terms of resolution and computational time.
We fix $Re=1$ (using the value of $U_0$ to obtain an {\it a priori} estimate of it), which is smaller than the critical 
value $\sqrt{2}$ of the Newtonian case, and we 
vary $Wi$ in a range of values larger than a critical one $Wi_c \approx 10$ 
corresponding to the onset of purely elastic instabilities \cite{BB10}. In all simulations the other 
parameters of the viscoelastic dynamics are $U_0=4$, $L=1/4$, $\nu_s=0.769$, $\eta=0.3$. 
The initial condition is obtained by adding a small random perturbation to the fixed point solution $\bm{u}^{(0)}$, $\bm{\sigma}^{(0)}$􏰁
and the system is evolved in time until a 􏰀statistically􏰁 steady state is reached.

Once the flow is in statistically stationary conditions, it is seeded with an ensemble of inertial particles, initially uniformly 
distributed in space and having randomly chosen velocities. 
Particle dynamics, Eqs. (\ref{eq:part_x}) and (\ref{eq:part_v}), are integrated by means of a standard Lagrangian approach 
using a second-order Runge-Kutta time-marching scheme; the velocity at particle positions is obtained by bilinear interpolation in space. 
Periodic boundary conditions are imposed on particle positions.  
A rather large number of Stokes time ($\tau_p$) values is examined, allowing to explore almost three decades in $St$ (for each considered 
flow, i.e. for each $Wi$). 
In the results reported in the following sections the number of particles is $N_p=10^4$ (tests with $N_p = 10^5$ did not show any 
appreciable difference on the statistics of single-particle observables).

\subsection{Elastic turbulence flows}\label{sec:3.1} 
\begin{figure}[htbp]
\centering
\subfloat[]{
\includegraphics[width=0.5\textwidth]{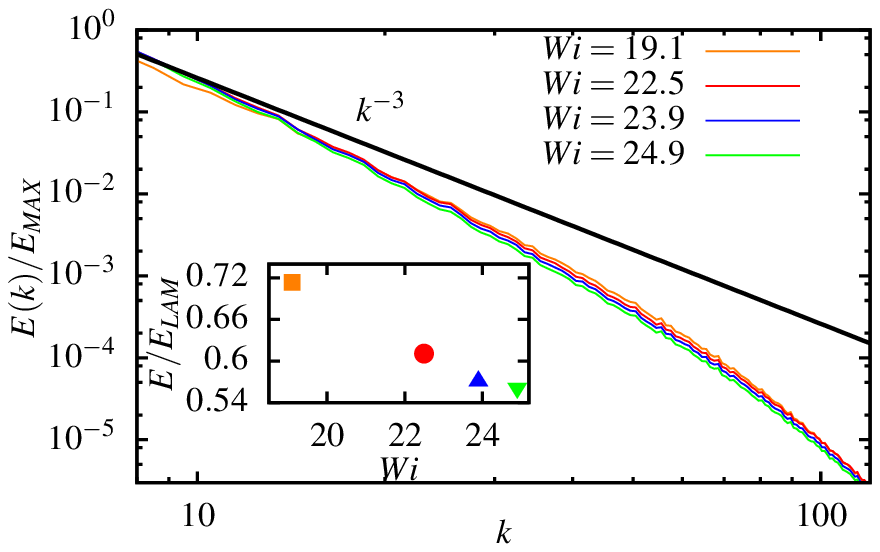}}

\subfloat[]{
\includegraphics[width=0.5\textwidth]{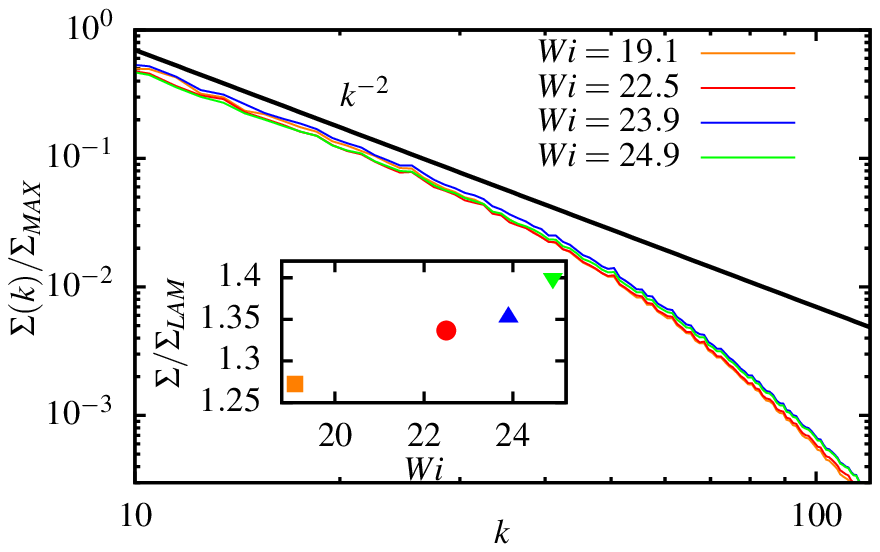}}
\caption{Time averaged spectra of kinetic energy $E(k)$ (a) and of the 
trace of the conformation tensor $\Sigma(k)$ (b), normalized by their maximum values, 
for different values of $Wi$ in the elastic turbulence regime.  
Inset of panel (a): temporally averaged kinetic energy $E=\overline{|\bm{u}|^2}/2$ normalized 
by its laminar value $E_{LAM}=U_0^2/4$. 
Inset of panel (b): temporally and spatially averaged square polymer elongation $\Sigma=\overline{tr(\bm{\sigma})}$ 
normalized by its laminar value $\Sigma_{LAM}=2+\frac{Wi^2}{4}$.}
 \label{fig:spectra}
\end{figure}
The transition from laminar to elastic turbulence states of the system specified by Eqs. (\ref{eq:oldroyd_kf_u}) and 
(\ref{eq:oldroyd_kf_s}) was previously studied in detail in \cite{BB10}. Here we are interested in working in the regime corresponding 
to Weissenberg numbers well above the threshold value $Wi_c$; the highest $Wi$ that we can safely reach in the present conditions 
is $Wi \approx 25$. For such values of $Wi$ the flow develops temporally and spatially irregular fluctuations associated with 
chaotic and mixing dynamics reminiscent of turbulence. From a statistical point of view, these turbulent-like features 
are described by the spectra of kinetic energy $E(k)$ and square polymer elongation or, equivalently, of the trace of the conformation 
tensor $\Sigma(k)$ (which is proportional to that of elastic energy). For both quantities we find power-law behaviors as 
$E(k) \sim k^{-\gamma}$ (Fig. \ref{fig:spectra}a) and $\Sigma(k) \sim k^{-\delta}$ (Fig. \ref{fig:spectra}b), indicating a whole range 
of active scales. The kinetic energy spectrum is characterized by an exponent $3.5<\gamma<3.6$ larger than $3$, pointing to smooth flow, 
and in reasonable agreement with the value measured in (three-dimensional) experiments (see, e.g., \cite{GS00}) 
and with theoretical predictions \cite{FL03} based on a simplified model corresponding to the large 
polymer elongation limit of Oldroyd-B model.
The spectral exponent of $\Sigma(k)$ is found to be $\delta \approx 2$, similarly to what is observed in numerical simulations 
of viscoelastic turbulence at higher $Re$ (and with finite extensibility models of polymer dynamics) \cite{DCP12,NDSBE16}.

In the insets of Fig. \ref{fig:spectra}a and Fig. \ref{fig:spectra}b, respectively, we report the behavior of global quantities, 
namely the (temporally averaged) kinetic energy $E=\overline{|\bm{u}|^2}/2$ and the (temporally averaged) trace of the conformation 
tensor $\Sigma=\overline{tr(\bm{\sigma})}$, normalized by their laminar values $E_{LAM}=U_0^2/4$ and $\Sigma_{LAM}=2+Wi^2/4$, 
as a function of the Weissenberg number. In agreement with previous observations \cite{BBBCM08,BB10}, we find that 
while the kinetic energy decreases with $Wi$, the square polymer elongation grows and this occurs faster than in laminar conditions. 
This suggests that polymers elongate by draining energy from the mean flow and, once sufficiently stretched 
they are capable of modifying the carrier flow through the term ${\frac{2\eta \nu_s}{\tau} \bm{\nabla} \cdot \bm{\sigma}}$ in 
Eq. (\ref{eq:oldroyd_kf_u}). The faster than laminar growth means that such elastic coupling is very efficient in sustaining the stretching 
of polymers.
 
\begin{figure*}[]
\subfloat{
\includegraphics[width=0.355\textwidth]{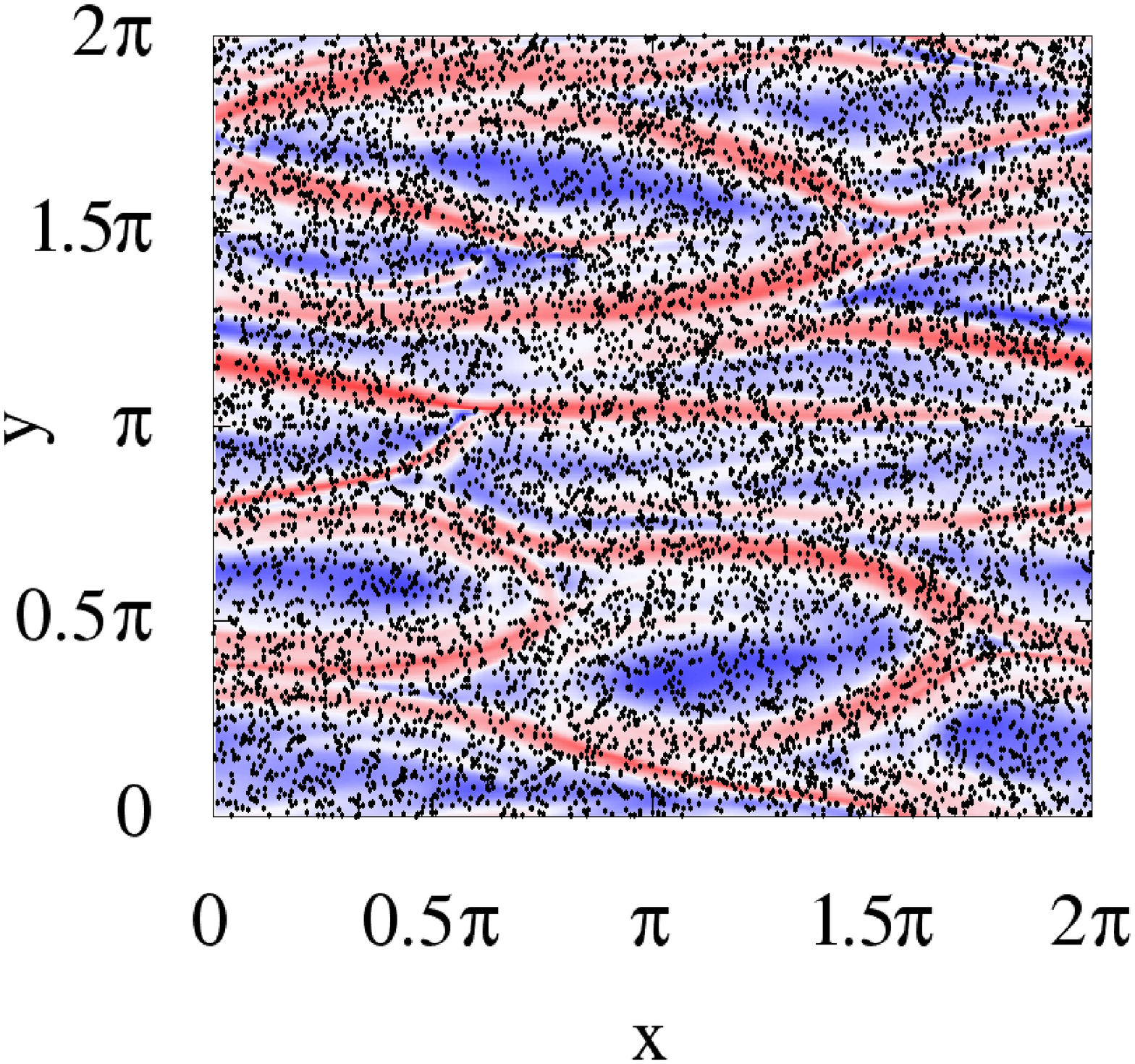}}
\subfloat{
\includegraphics[width=0.2825\textwidth]{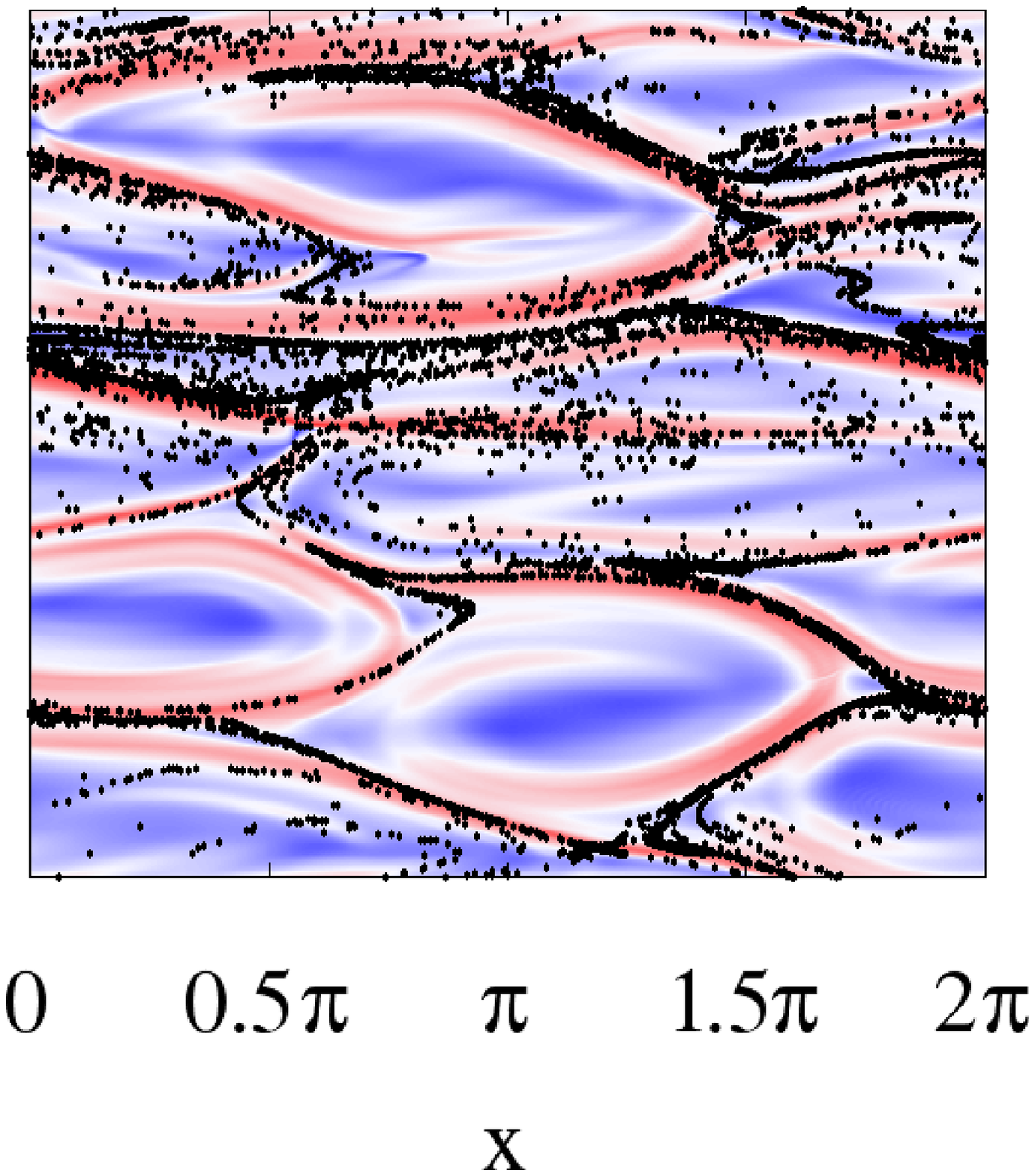}}
\subfloat{
\includegraphics[width=0.355\textwidth]{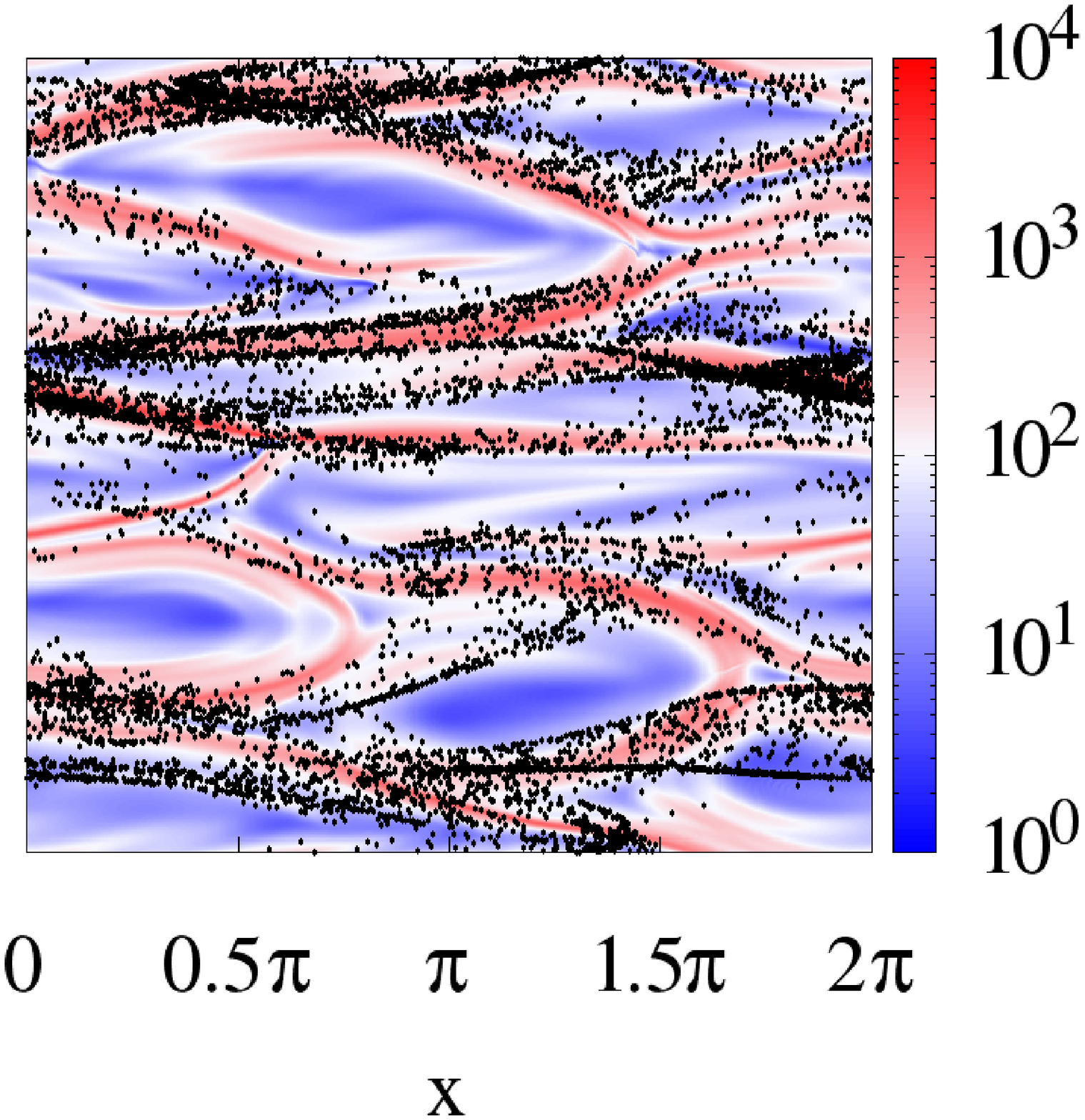}}

\subfloat{
\includegraphics[width=0.355\textwidth]{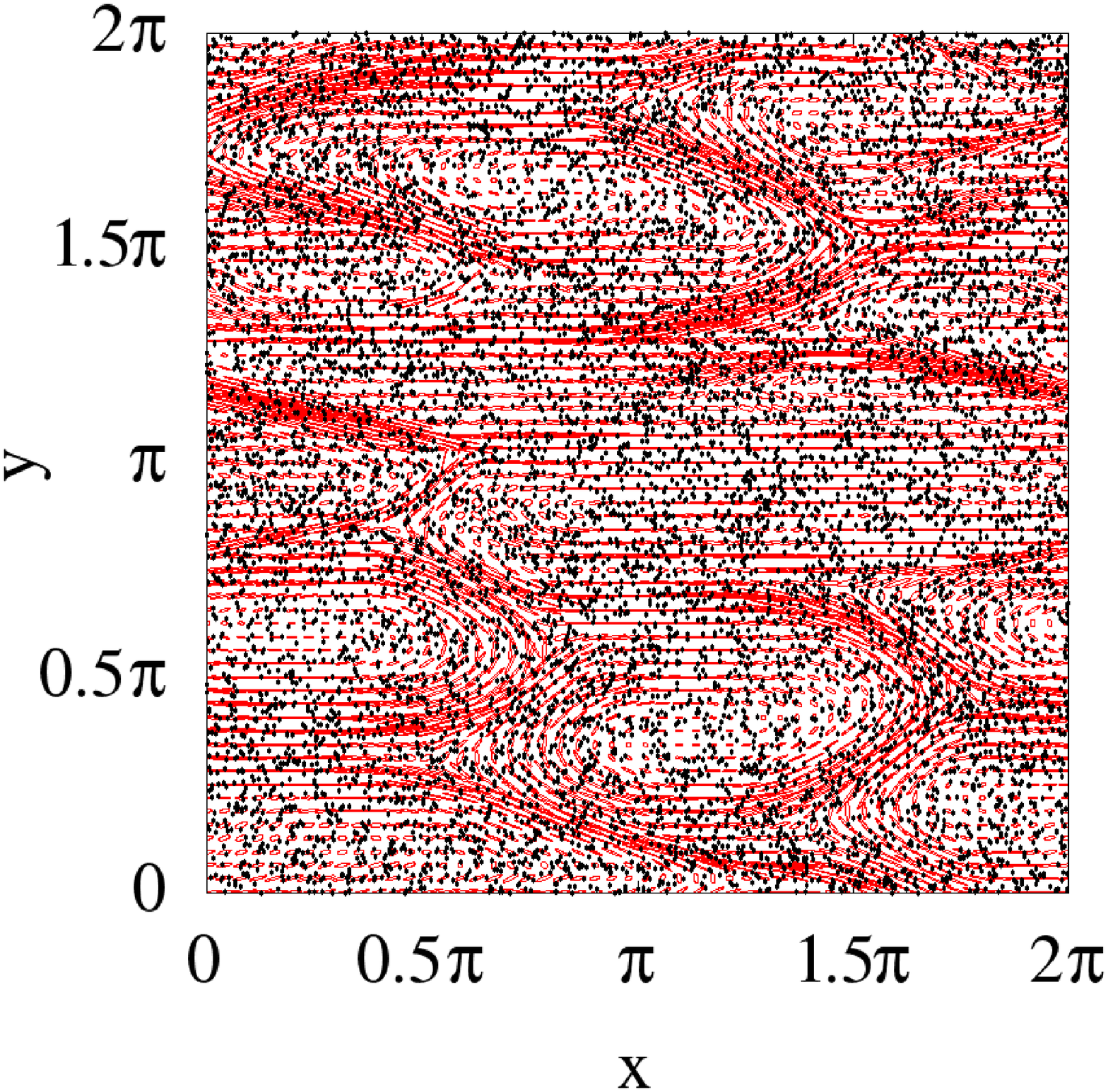}}
\subfloat{
\includegraphics[width=0.3\textwidth]{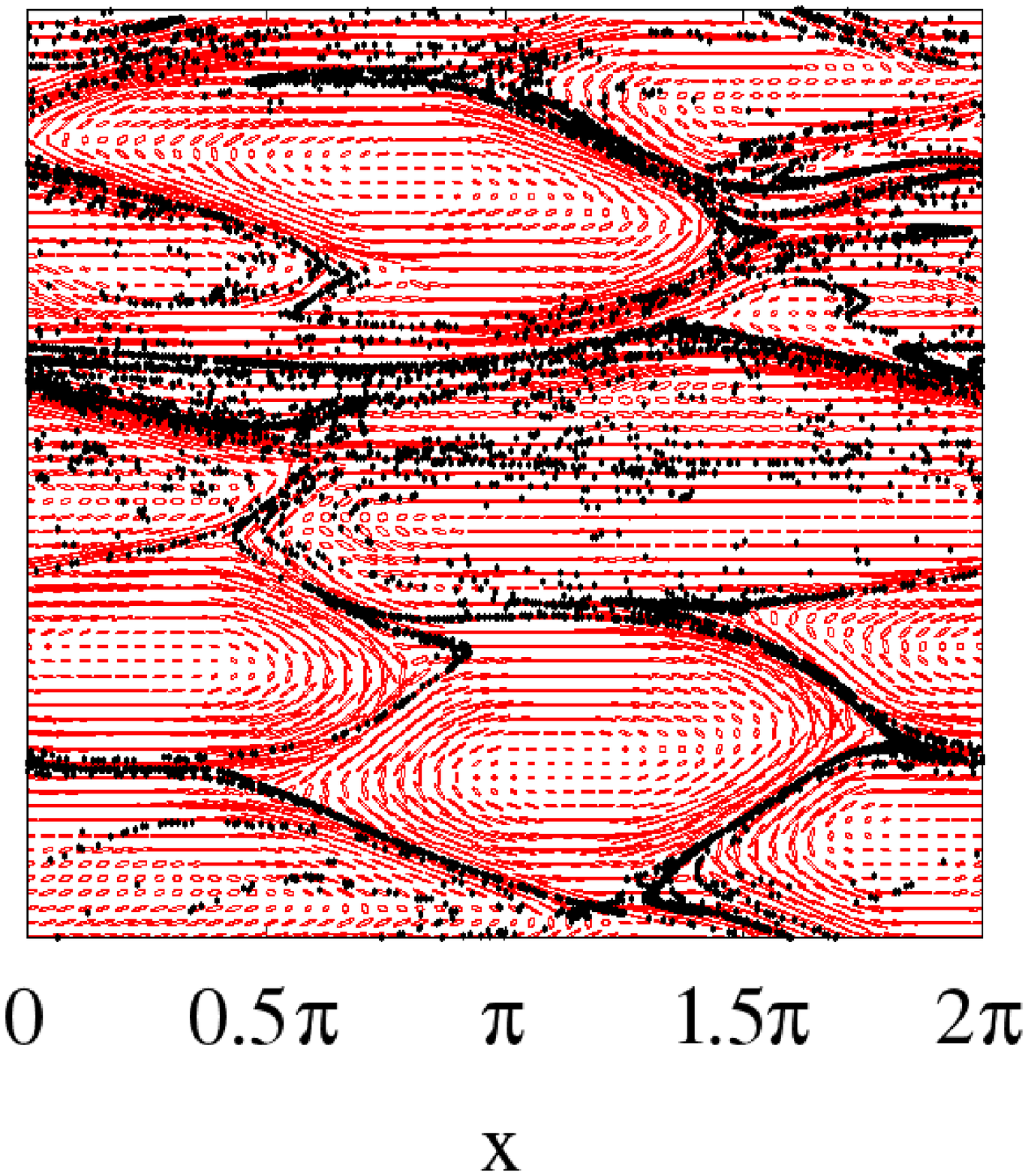}}
\subfloat{
\includegraphics[width=.3\textwidth]{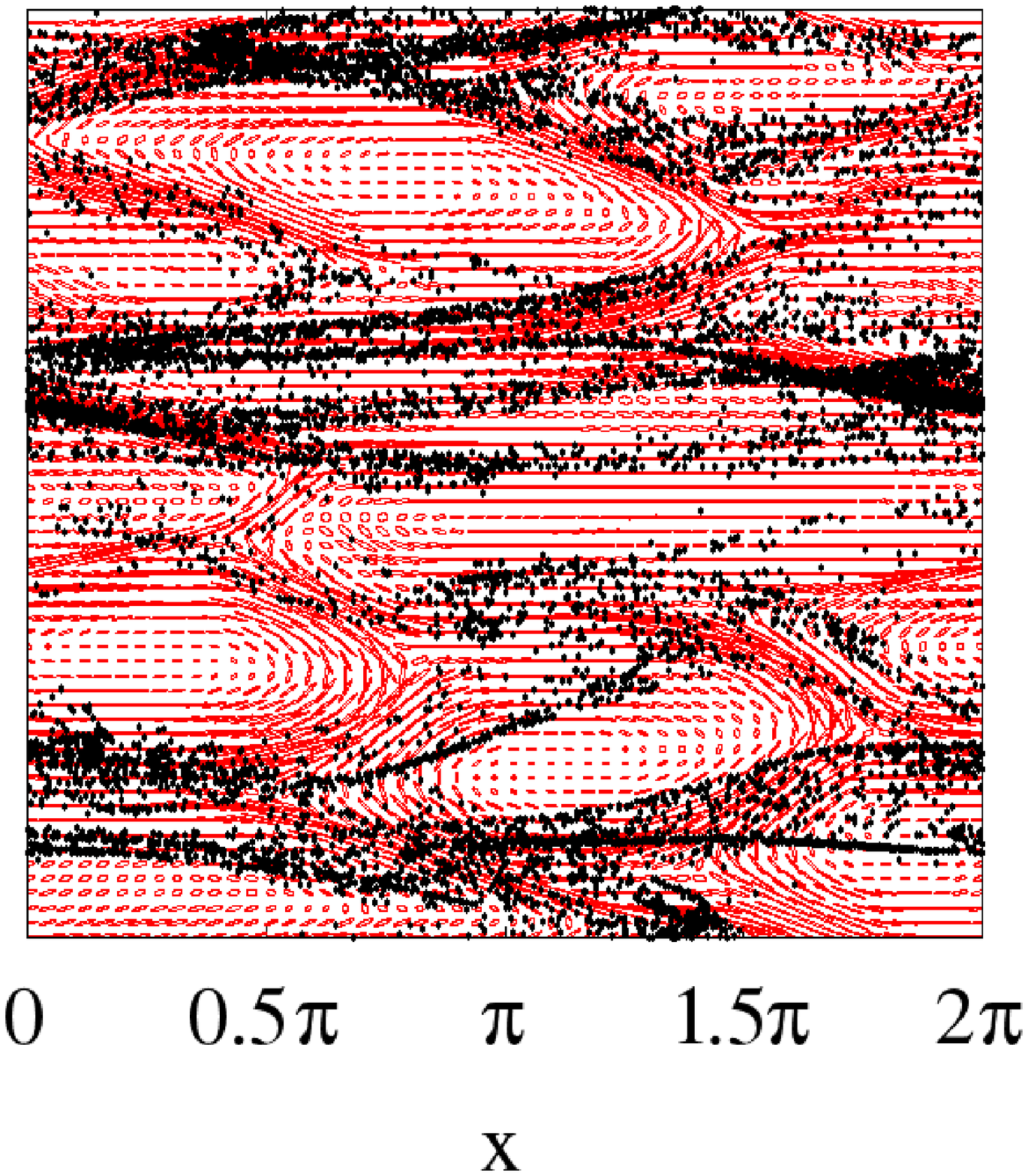}}

\subfloat{
\includegraphics[width=0.355\textwidth]{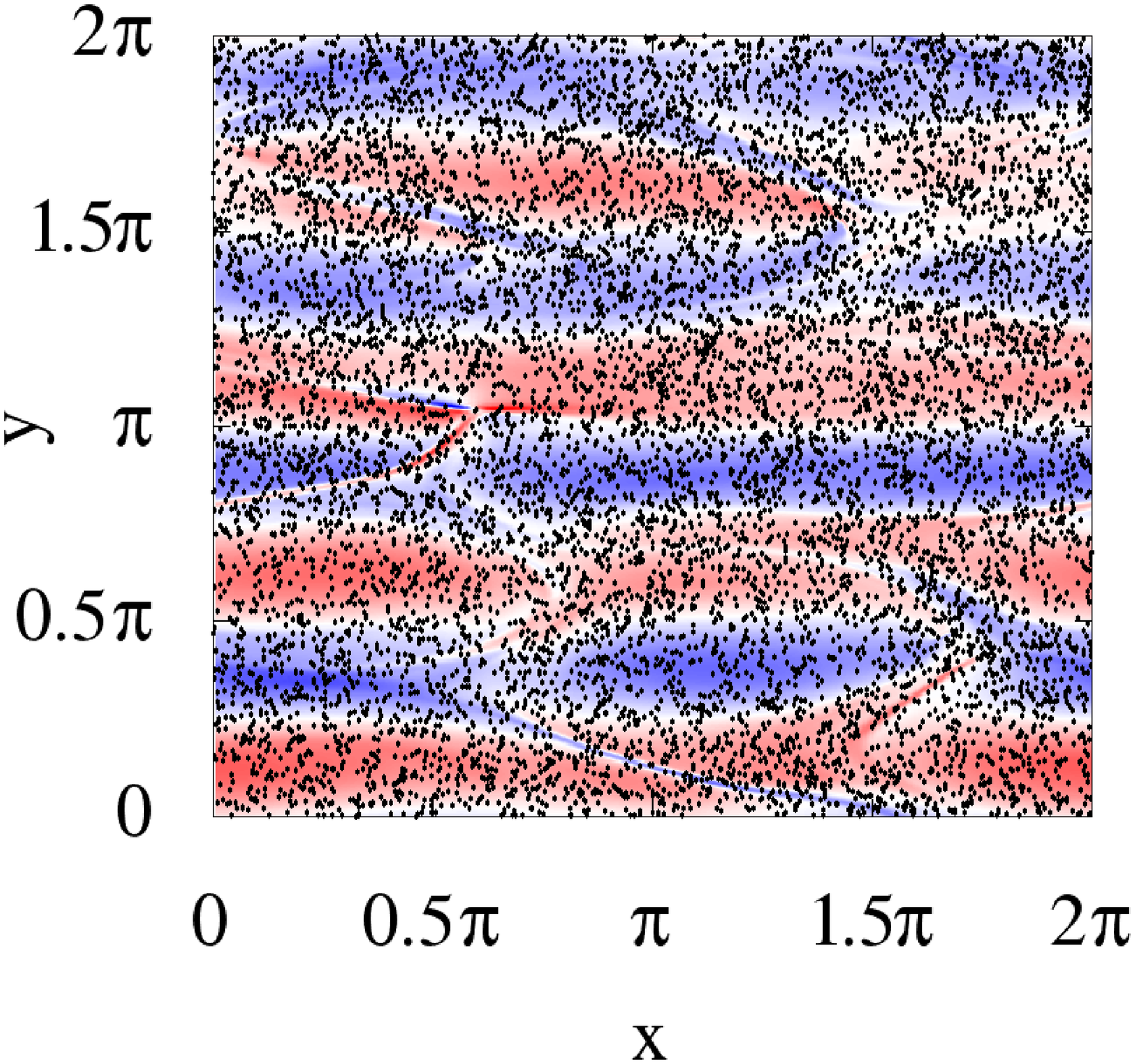}}
\subfloat{
\includegraphics[width=0.285\textwidth]{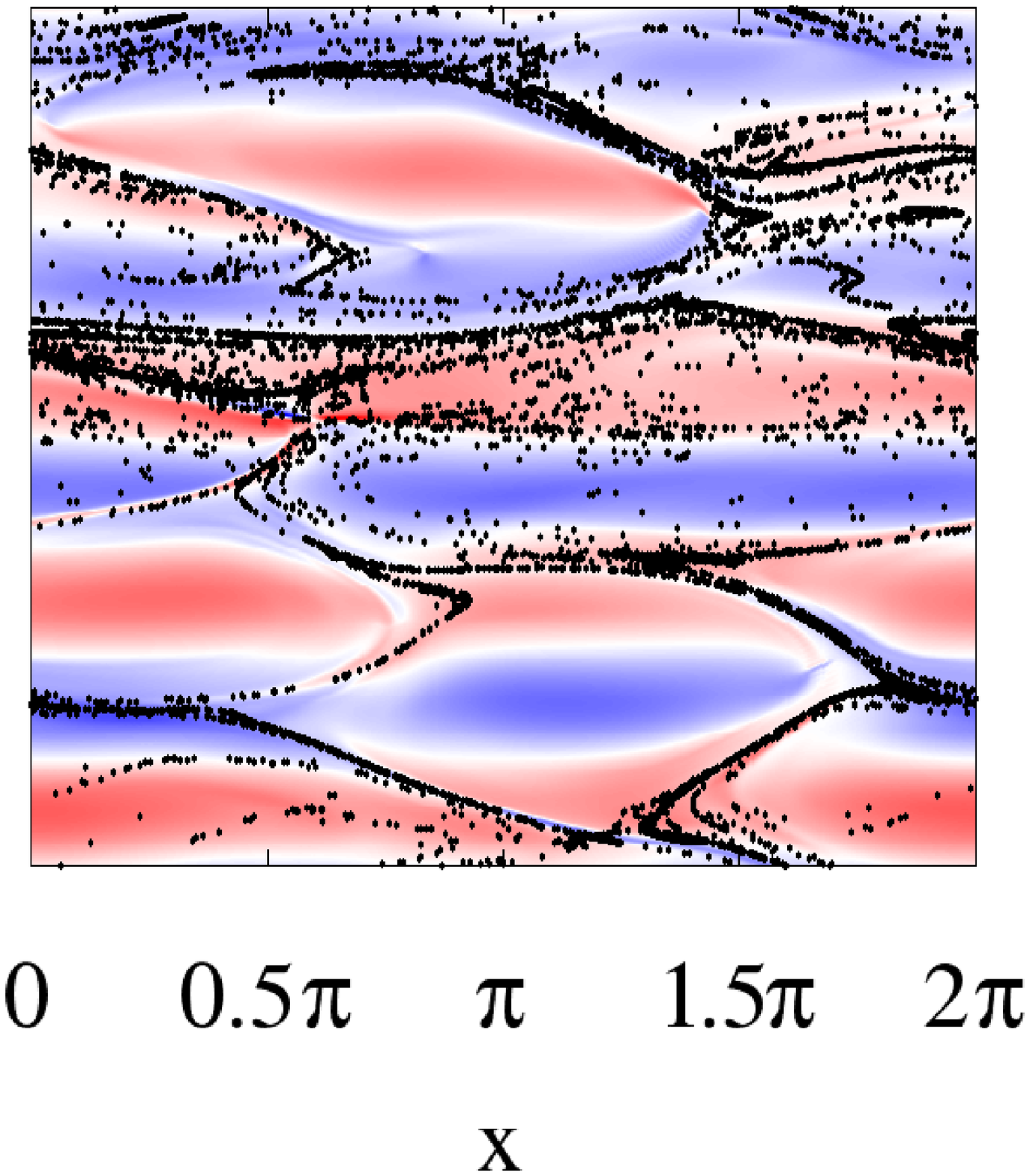}}
\subfloat{
\includegraphics[width=0.355\textwidth]{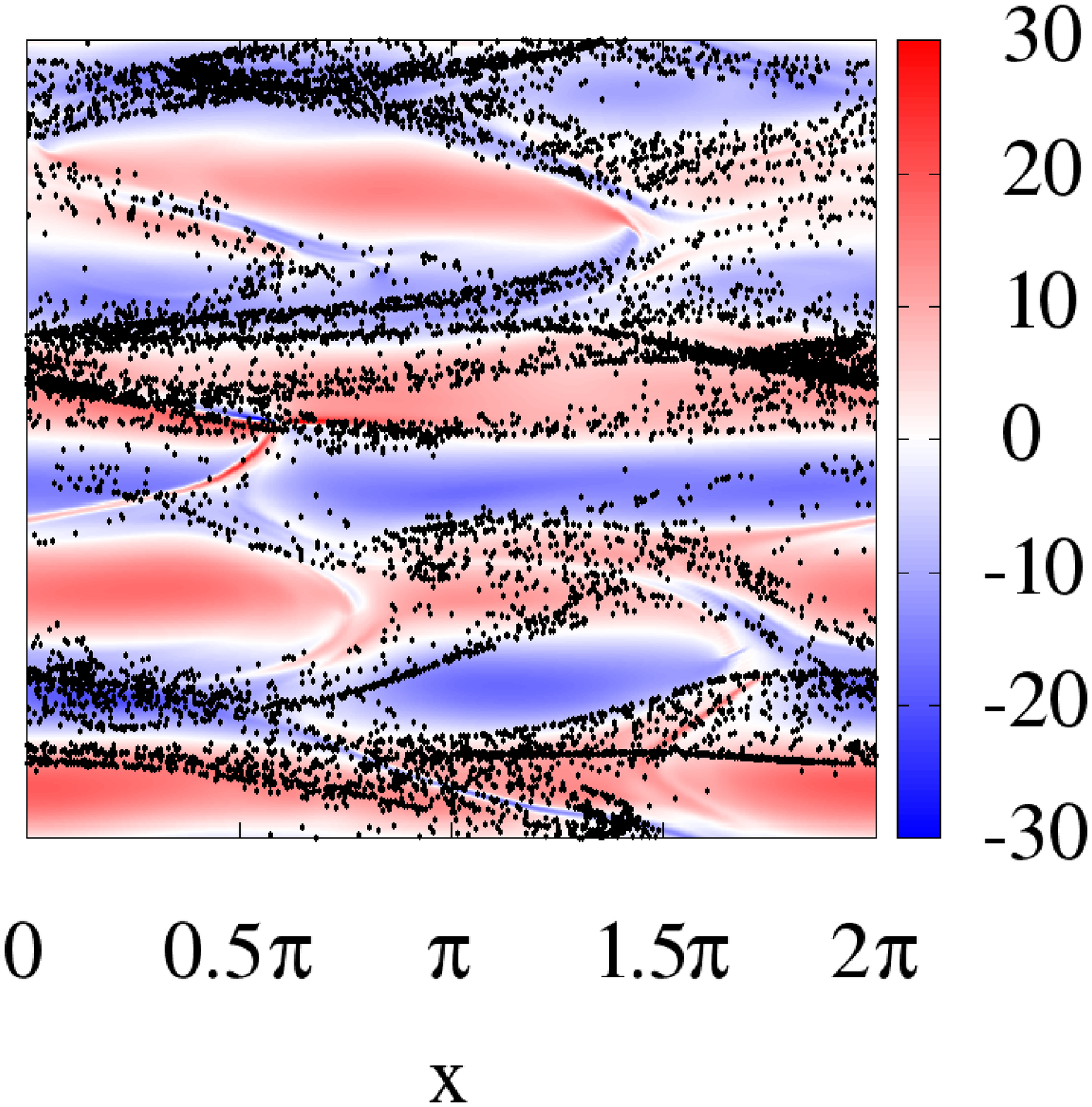}}
\caption{Particle distribution (black dots) for $St=0.016$, $St=0.657$ and $St=5.75$ (from left to right) 
at a fixed instant of time in statistically stationary conditions at $Wi=23.9$ and $Re=0.664$; the number of 
particles is $N_p=10^4$. The pseudocolor plots 
in the upper and bottom panels respectively correspond to instantaneous snapshots of $[tr(\bm{\sigma})](x,y)$ 
and vorticity $\zeta(x,y)=\bm{\nabla} \times \bm{u}(x,y)$ at the same time for which particles are plotted. 
In the central panel, particles are plotted together with 
an ellipsoid-glyph visualization of the polymer conformation tensor $\bm{\sigma}$.}
\label{fig:partdistr_fields}
\end{figure*}

\subsection{Preferential concentration effects}\label{sec:3.2}
We now discuss particle dynamics, starting from an analysis of the statistical properties of their spatial distribution 
in relation with the main dynamical features of the viscoelastic fluid flow. 
Throughout all this study $\tau_{\dot{\gamma}} \approx 0.1$ and the polymer relaxation time $\tau$ is typically 
larger than both $\tau_{\dot{\gamma}}$ and $\tau_p$. 
As it is evident from Fig. \ref{fig:partdistr_fields} (where $Wi=23.9$ and $St$ increases from left to right), 
due to their inertia, particles non-homogeneously distribute in space. Let us remark, here, that Lagrangian tracers 
(i.e. non-inertial particles for which $St=0$) evolve according to Eq. (\ref{eq:part_x}) only and, consequently, 
homogeneously sample the flow field, if initially uniformly seeded in it as in this case.    
In the presence of inertia, the non-homogeneous character of the particle distribution appears to vary 
non-monotonically with $St$, with a maximum for intermediate values of this parameter. This is in agreement 
with intuitive expectations: for very small $St$ one should recover tracer dynamics, while for very large $St$ particle 
dynamics should be essentially insensitive to the flow. 
In Fig. \ref{fig:partdistr_fields} both small-scale inhomogeneities and larger scale modulations of the particle 
distributions are seen. A striking feature is, however, the accumulation of particles along thin filamentary structures 
characterized by large polymer elongations, i.e. large values of $tr(\bm{\sigma})$ (see upper panel of 
Fig. \ref{fig:partdistr_fields}). 
Such highly elastic filaments, propagating along the mean flow direction, are associated with the stretching of polymers by 
the largest gradients of the mean velocity field \cite{BB10}. Similar wavy patterns also characterize the vorticity field 
$\zeta=\bm{\nabla} \times \bm{u}$ (see bottom  panel of Fig. \ref{fig:partdistr_fields}), due to the coupling between polymeric 
and velocity dynamics. 
The strong correlation between the spatial organization of the particle distribution 
and that of the polymer conformation tensor field is further evidenced by plotting the latter by means of an ellipsoid 
representation of the local (in space) principal elongations (central line of Fig. \ref{fig:partdistr_fields}).

\begin{figure*}[t]
\centering
\includegraphics[width=0.8\textwidth]{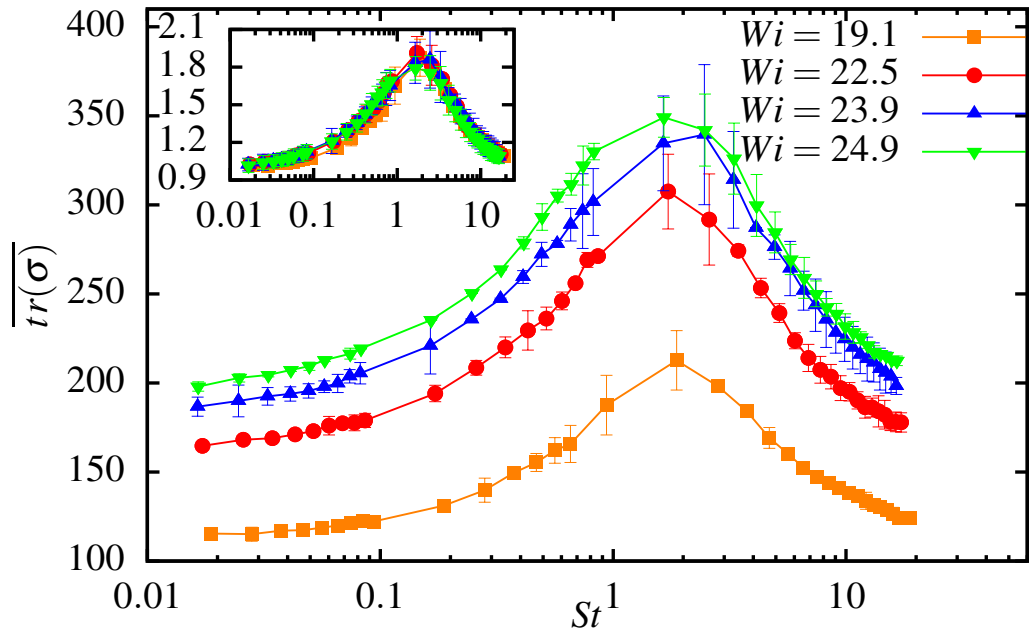}
\caption{Average trace of the conformation tensor $\overline{tr(\bm{\sigma})}$ experienced by particles as a function of $St$ and 
for different Weissenberg numbers. Here temporal averages are performed over $50$ snapshots of $tr(\bm{\sigma})$
(and simultaneous particle distributions) corresponding to different instants of time separated by an interval larger 
than the typical flow time scale. 
The inset shows the same after rescaling $\overline{tr(\bm{\sigma})}$ with its value computed using Lagrangian tracers 
$\overline{tr(\bm{\sigma})}_{St=0}$.}
\label{fig:part_trace}
\end{figure*}

In order to quantitatively assess this point, we computed the trace of the conformation tensor 
$\overline{tr(\bm{\sigma})}$, averaged over the whole space domain and a long time history, experienced by inertial particles 
as a function of Stokes and for different values of $Wi$. The curves reported in Fig. ~\ref{fig:part_trace} have 
non-monotonic behavior, with a maximum of $\overline{tr(\bm{\sigma})}$ for $St \approx 1$. 
Their qualitative features are generic with respect to the value of the Weissenberg number. 
Indeed, as shown in the inset of Fig. \ref{fig:part_trace}, after rescaling $\overline{tr(\bm{\sigma})}$ with 
the same quantity $\overline{tr(\bm{\sigma})}_{St=0}$ computed for tracers in the same flow (for each $Wi$) 
we obtain a good collapse of the data, indicating the $Wi$ independence of this observable.
These results demonstrate that when inertia is increased, and not too large, particles have an increasing tendency 
to concentrate where polymers are highly stretched. Moreover, as it is clear from the inset of the figure, independently of $St$, 
inertial particles experience larger values of  $tr(\bm{\sigma})$ than fluid-flow Lagrangian tracers. 
  
To understand the phenomenology described above, one has to relate elastic filaments to the velocity field that transports particles. 
A hint in this sense comes from inspection of ellipsoid-glyph visualizations of the polymer conformation tensor 
(Fig. \ref{fig:partdistr_fields}). 
In these plots, the presence of regions of recirculating motion is more evident, with elastic filamentary stuctures playing the role 
of flow separatrices (as also observed in numerical simulations of viscoelastic cellular flows \cite{GP17}). 
Some details on the formation of vortices in this elastic turbulence flow can be found in \cite{BB10}. 
Here, instead, we want to focus on their impact on particle dynamics. 
In fact, several previous studies (see, e.g., \cite{maxey-1987,squires1991preferential,bec2006acceleration}) have demonstrated that 
small and heavy inertial particles migrate to strain dominated flow regions because they are expelled from vortical regions 
by centrifugal forces. At least in the small $St$ limit, this can be explained as follows. 
From a Taylor expansion of Eq. (\ref{eq:part_v}) at first order in $\tau_p$ one has 
$\bm{v} \simeq \bm{u}-\tau_p(\partial_t \bm{u}+\bm{u}\cdot\bm{\nabla}\bm{u})$ \cite{maxey-1987}. 
Then, for the divergence of the particle velocity one obtains 
\begin{equation}
\bm{\nabla} \cdot\bm{v} = -\tau_p \; tr \left[ \bm{\nabla} \bm{u} \cdot \left(\bm{\nabla} \bm{u}\right)^T \right],   
\label{eq:div_v}
\end{equation}
using the incompressibility of the velocity field $\bm{u}$. 
Decomposing the fluid velocity gradient $\bm{\nabla}\bm{u}$ into its symmetric $\bm{S}$ and anti-symmetric part $\bm{\Omega}$, we then have 
\begin{equation}
\bm{\nabla} \cdot\bm{v} = 2 \tau_p Q, 
\label{eq:div_v_Q}
\end{equation}
where, up to a prefactor redefinition, 
\begin{equation}
Q = \frac{1}{2}(\Omega_{ij}\Omega_{ij}-S_{ij}S_{ij})
\label{eq:ow_param}
\end{equation}
is Okubo-Weiss parameter \cite{okubo1970horizontal,weiss1991dynamics}. 
In the above equation $S_{ij}$ and $\Omega_{ij}$ respectively indicate the elements of the rate-of-strain ($\bm{S}$) and 
rate-of-rotation ($\bm{\Omega}$) tensors, and summation over repeated indices is assumed.
Particles concentrate due to (weak) compressibility of their velocity, that is where $\bm{\nabla} \cdot \bm{v}<0$. From Eq. (\ref{eq:div_v_Q}) 
it is seen that this condition translates into negative values of $Q$, meaning that particles are expected to preferentially sample 
strain dominated regions (using Eq.(\ref{eq:ow_param})). 

\begin{figure*}[hbtp]
\centering
\includegraphics[width=0.8\textwidth]{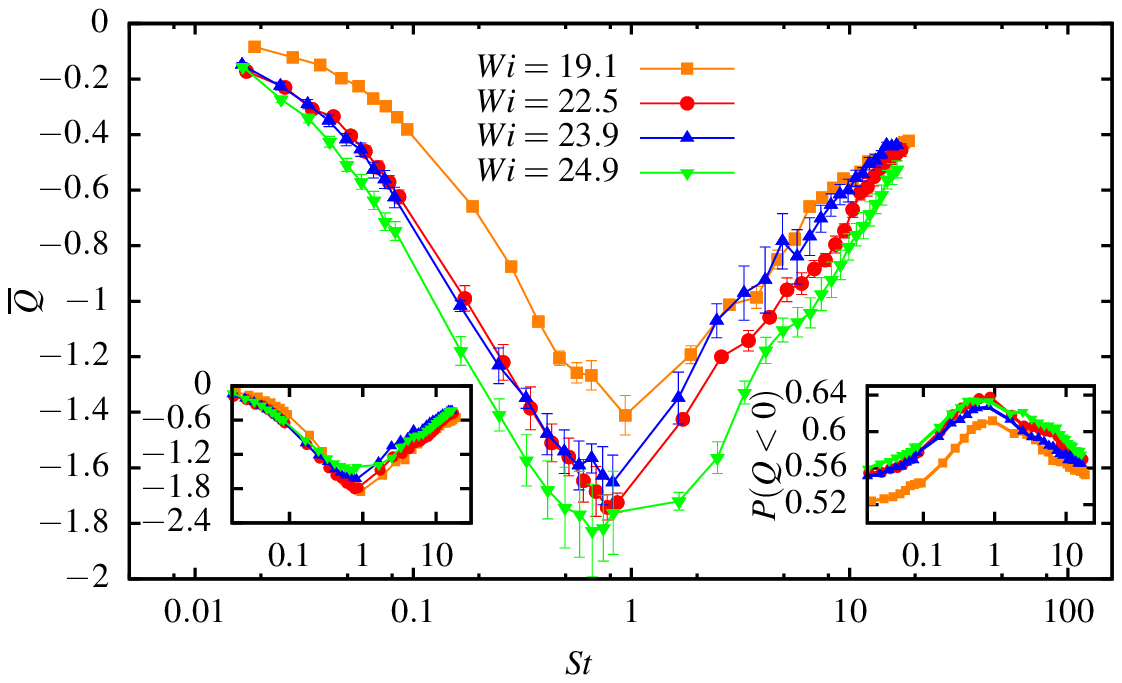}
\caption{Average Okubo-Weiss parameter $\overline{Q}$ experienced by particles as a function of $St$. 
Here temporal averages are performed over $50$ snapshots of $Q$ (and simultaneous particle distributions) corresponding to 
different instants of time separated by an interval larger than the typical flow time scale.
Left inset: normalized Okubo-Weiss parameter $\overline{Q}/{Q^{rms}_{St=0}}$ as a function of $St$, where $Q^{rms}_{St=0}$ is the 
root-mean-square value of $Q$ experienced by Lagrangian tracers. 
Right inset: probability that a particle is in strain dominated regions ${P}(Q < 0)$ as a function of Stokes. 
} 
\label{fig:ow_part}
\end{figure*}

Figure \ref{fig:ow_part} shows the spatially and temporally averaged Okubo-Weiss parameter measured at particle positions versus $St$ and 
for different Weissenberg numbers. The results support the above argument and provide a quantitative confirmation of what observed 
from Fig. \ref{fig:partdistr_fields}. Indeed, $\overline{Q}$ is found to be always negative, which suggests that particles are ejected 
from recirculating regions to get more concentrated in regions dominated by strain, where polymers are highly elongated. 
Also in this case, the effect is maximum (i.e. $\overline{Q}$ is minimum) for $St \approx 1$. The effect of varying $Wi$  
is found to be quite weak. In the left inset of Fig. \ref{fig:ow_part} we show the behavior, versus $St$, of Okubo-Weiss parameter  
rescaled with its root-mean-square (rms) value computed for tracers $Q^{rms}_{St=0}$ (since $\overline{Q} = 0$ 
for Lagrangian tracers and, equivalently, for the Eulerian fluid flow, from numerical simulations). After rescaling, 
the results are only very weakly dependent on $Wi$.
We end this section by commenting on the right inset of Fig. \ref{fig:ow_part}. The plot presents the probability $P(Q < 0)$ that 
a particle is in a strain dominated region, which is computed as the ratio between the number of particles at positions where 
$Q<0$ and the total number of particles, as a function of $St$. 
The probability $P(Q<0)$ generally takes values larger than the one realized in the limit of very small $St$.  
Despite not large, such an increase of $P(Q<0)$ indicates that inertial particles are more concentrated than tracers in 
regions where $Q<0$. 
Finally, we observe that the effect is, again, maximum for $St \approx 1$ and weakly dependent on $Wi$. 

\subsection{Correlation dimension of small-scale clusters}\label{sec:3.3}
The previous analysis allowed us to reveal some relations between the inhomogeneities of the particle distribution and 
flow structures. The fine scale properties of particle clustering are, however, a more general consequence of the contraction 
of volumes in the phase space of the dissipative system of Eqs. (\ref{eq:part_x}) and (\ref{eq:part_v}) \cite{bec2003fractal}. 
In both laminar unsteady and turbulent flows, it has been shown that the motion of inertial particles at small scales is highly 
non-trivial and, at sufficiently large times, it occurs on a fractal set \cite{bec2003fractal,GM16}. 
A possibility to quantitatively characterize clustering is then to measure the fractal dimension, in physical space, of the attractor 
of the dynamics. When this is smaller than the dimension of the full physical space, particle pairs are more likely separated 
by small distances. Within this framework, a common indicator is the correlation dimension $D_2$ \cite{GP1983}, 
which is defined as:
\begin{equation}
 D_2 = \lim_{r\rightarrow 0} \frac{\log[C(r)]}{\log(r)},  
\label{eq:D2}
\end{equation}
with the correlation sum $C(r)$ given by
\begin{equation}
C(r)=\lim_{N_p\to \infty }\frac{2}{N_p(N_p-1)}\sum_{i,j>i}^{N_p}\Theta (r-\left | {\bm{x}_i}-{\bm{x}_j} \right |) \nonumber,
\end{equation}
where $\Theta$ is the Heaviside step function and $\bm{x}_i$ and $\bm{x}_j$ are the positions of particles belonging to pair $(i,j)$. 
Equation (\ref{eq:D2}) then means that, for small $r$, the probability to find particle pairs separated by a distance less than $r$ scales 
as $C(r) \sim r^{D_2}$. 
\begin{figure*}[htbp]
\centering
\includegraphics[width=0.8\textwidth]{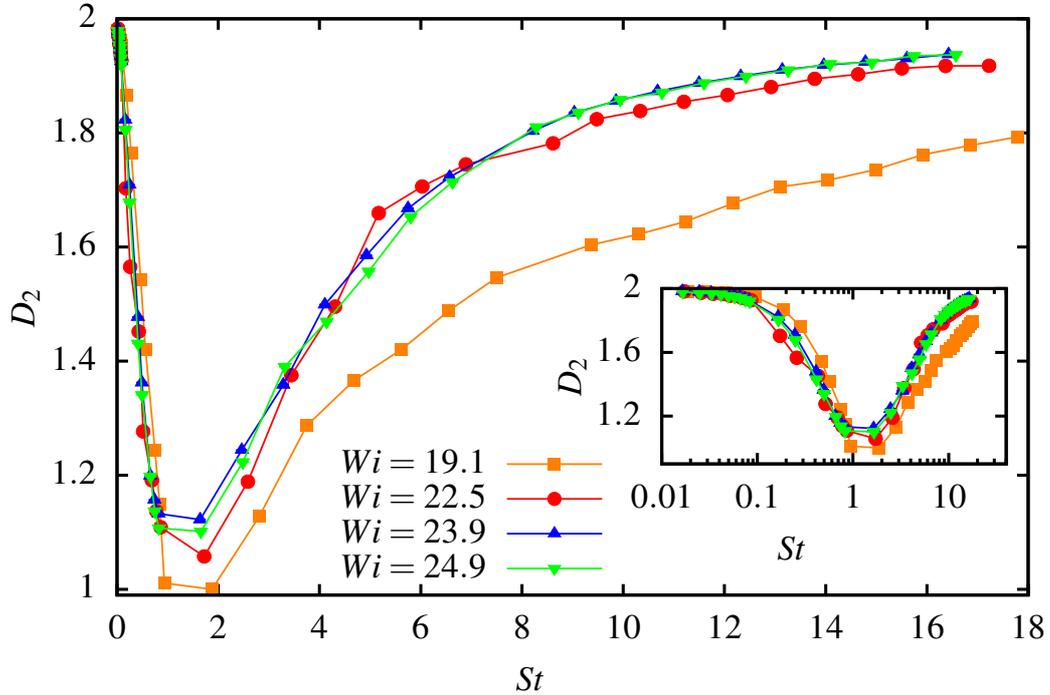}\hspace{2pc}%
\caption{Correlation dimension  $D_2$ of particle distributions as a function of $St$ and for different Weissenberg numbers; 
uncertainties are of the order of point size. Small-scale clustering is maximum for $St \sim 1$, where $D_2$ has a minimum. 
The relative difference between the values obtained for $Wi=19.1$ and the corresponding ones for $Wi=24.9$ is not larger than $0.15$.
The inset shows the same in log-linear scale.} 
\label{fig:D2}
\end{figure*}

The behavior of $D_2$ as a function of the Stokes number for different values of $Wi$ is presented in Fig. \ref{fig:D2}. 
It is seen that the correlation dimension decreases from a value, which is realized in the limit of very small $St$, close to $D_2=2$ 
(corresponding to tracers homogeneously filling the whole space domain) to attain a minimum value of $D_2 \approx 1$ 
for $St \approx 1$. For even larger values of the Stokes number, $D_2$ grows to approach again the space filling value 
of $2$ (expected for large inertia particles that are insensitive to the flow) in the limit of very large $St$. 
We find that the correlation dimension is weakly dependent on the Weissenberg number, for the values of $Wi$ explored here. 
The maximum relative difference, for fixed $St$, is found to not exceed $0.15$. 
We can therefore conclude that small-scale clustering is a generic and quite effective phenomenon in elastic turbulence flows, 
producing, at its maximum, particle accumulation on quasi one-dimensional fractal sets.  
Our results are qualitatively similar to previous ones obtained in simulations of 2D smooth random flows \cite{bec_2005}. 

\subsection{Elastically driven turbophoresis}\label{sec:3.4}
\begin{figure}[htbp]
\centering
\vspace{-1em}\hspace{-1.5em}\subfloat[]{
\includegraphics[scale=1.1]{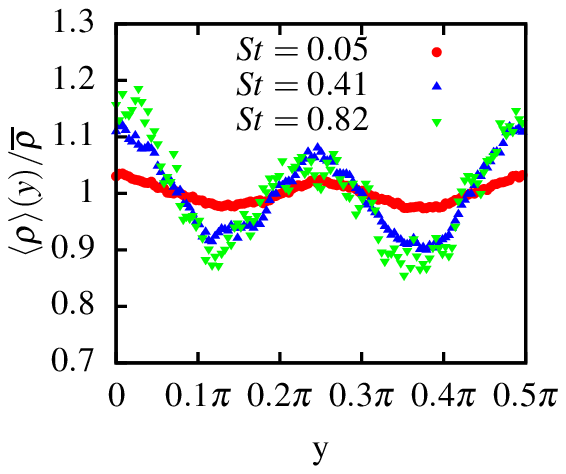}} 
 
\vspace{-2em}\subfloat[]{
\includegraphics[scale=1.1]{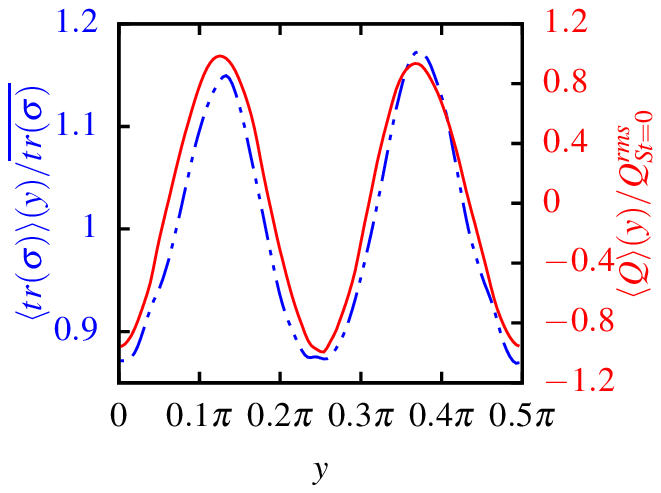}} 

\vspace{-2em}\subfloat[]{
\includegraphics[scale=1.1]{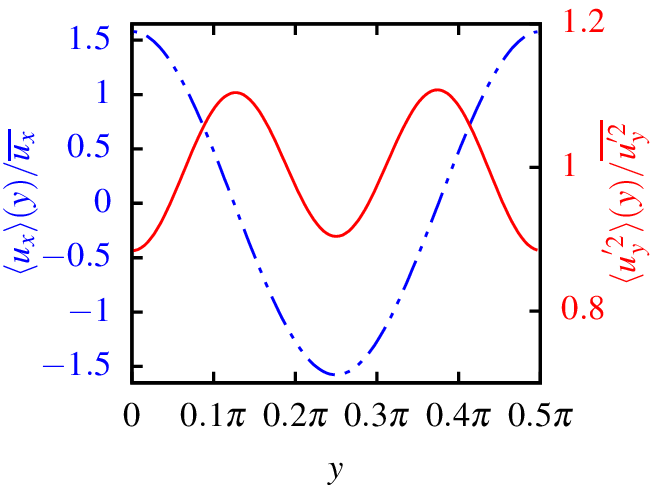}}
\caption{(a) Particle number density profiles $\lra{\rho}(y)/\overline{\rho}$, normalized by the 
global mean uniform density ($\overline{\rho}=1/L_0$), for three different Stokes numbers. 
(b) Normalized profile of the trace of the conformation tensor $\lra{tr(\bm{\sigma})}(y)/\overline{tr(\bm{\sigma})}$ 
(blue dashed line, left axis) and of Okubo-Weiss parameter $\lra{Q}(y)/Q^{rms}_{St=0}$ (red solid line, right axis). 
(c) Normalized profiles of the longitudinal velocity $\lra{u_x}(y)/{\overline{u_x}}$, where $\overline{u_x}=U$, 
(dashed blue line, left axis) and of the fluctuations of the shear-normal kinetic energy $\lra{u'^2_y}(y)/\overline{u'^2_y}$ 
(red solid line, right axis) of the fluid flow. The plots in (a-c) refer to statistically stationary conditions for $Wi=23.9, Re=0.664$. 
All the shown profiles are further averaged over the wavelength defining the periodicity of the mean flow $\ell=LL_0=\pi/2$. 
}
\label{fig:rho-trace-uysqr}
\end{figure}

In this section we investigate the large-scale properties of the particle spatial distribution. Here, a motivation 
is provided by the observation that some form of modulation along the mean-shear direction ($y$)
is already apparent from the visualizations of Fig. \ref{fig:partdistr_fields}. 
To analyze how this is related to the flow features, we introduce the particle number density field $\rho(\bm{x},t)$ 
and focus on the profiles along the direction of inhomogeneity $y$ of both $\rho$ and flow statistics.  
For each considered quantity, the $y$-profile is obtained by averaging over the mean-flow direction $x$ and time, 
which leaves a function of $y$ only. We indicate profiles with 
$\lra{(...)}$. Note that this type of average is related to the global one introduced in Sec. \ref{sec:2} by
\begin{equation}
\overline{(...)} = \frac{1}{L_0}\int_{0}^{L_0} \lra{(...)} dy.
\nonumber
\end{equation}

Figure \ref{fig:rho-trace-uysqr} presents the profiles of $\rho$ (panel (a)) for three different Stokes numbers,
as well as those of several flow related quantities (panels (b) and (c)), in a state of elastic turbulence 
(with $Wi=23.9$ and $Re=0.664$). 
All profiles are normalized by their, uniform, global average value to stress the deviations from it. 
We remark that $\lra{u_y}(y)=0$ with very good accuracy in the numerics, as expected from symmetry considerations.  
We also note that the shown results are obtained by further averaging them over one forcing wavelength $\ell=LL_0=\pi/2$.  
Comparing panels (a) and (b) of the figure, we see that, consistently with the previous analysis, particles are maximally 
concentrated where the longitudinally averaged Okubo-Weiss parameter $\lra{Q}(y)$ is minimum. Remark that here $\lra{Q}(y)$ 
is normalized by $Q^{rms}_{St=0}$ due to the fact that $\overline{Q}_{St=0}=0$. 
Nevertheless, in such regions of minimal $\lra{Q}(y)$, the profile of the trace of the conformation tensor 
$\lra{tr(\bm{\sigma})}(y)$ is now found to be minimum too. This apparently contradicts the observation made in Sec. \ref{sec:3.2} 
that particles aggregate in regions of highly elongated polymers. This contradiction is 
solved by considering that profiles result from a spatial averaging procedure. 
Indeed, all information about the spatial structure along the longitudinal direction is lost 
in them, which are functions of the transversal direction only. This particularly applies to 
the information about the extent of vortices along $x$, from which particles are expelled, and about the orientation, 
with respect to $x$, of the separatrices, by which particles tend to be attracted and that colocate with high polymer elongation regions. 
While the profile $\lra{Q}(y)$ receives contributions only from the transverse fluctuating component of the 
velocity field, indeed $\lra{Q}(y) =  -\partial_y^2  \lra{u_y'^2}(y)$ (with prime indicating the fluctuation), 
the trace of the conformation tensor is dominated by the contribution of the mean flow, $\lra{u_x} (y)$, 
to polymer stretching and $\lra{tr(\sigma)}(y) /\overline{tr(\sigma)} \simeq \lra{\sigma_{11}}(y) / \overline{\sigma_{11}}$.

From the above discussion it should be clear that the large-scale inhomogeneities of $\rho$ cannot be explained directly 
in terms of the averaged profiles. In fact, they are a manifestation of the turbophoresis phenomenon. 
In a nutshell, this corresponds to the migration of inertial particles 
from regions of high to regions of low eddy diffusivity that occurs in turbulent flows with non-homogeneous mean flow. 
Turbophoresis has been mainly studied in wall-bounded flows, because of their relevance for industrial and 
environmental applications related to particle deposition \cite{caporaloni1975,bkhm1992,picano2009spatial}.  
Interestingly, using the three-dimensional (3D) Newtonian turbulent Kolmogorov flow, it was recently shown that 
turbophoretic segregation is independent of the presence of walls \cite{DCMB16}. 
Also in that case particles accumulate in regions of minimum turbulent diffusivity, but the  
spatial distribution of the latter with respect to the mean flow differs from the one found 
in geometrically confined flows. 
\begin{figure}[]
\centering
\vspace{-1em}\subfloat[]{
\includegraphics[scale=1]{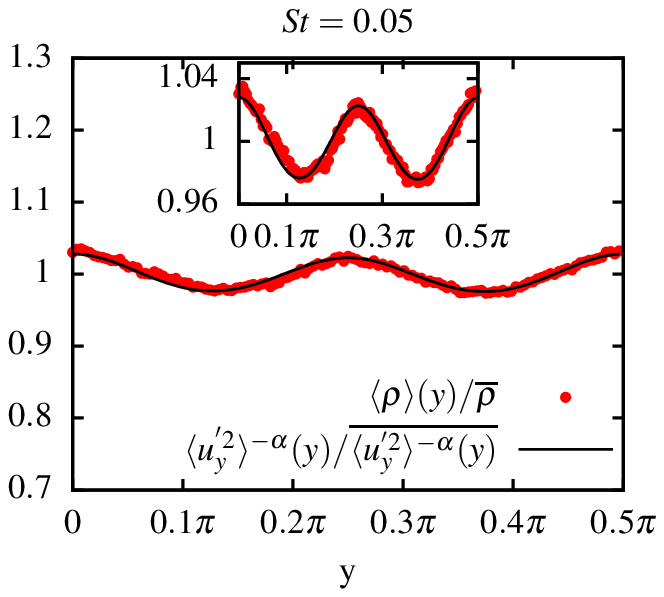}}

\vspace{-1.6em}\subfloat[]{
\includegraphics[scale=1]{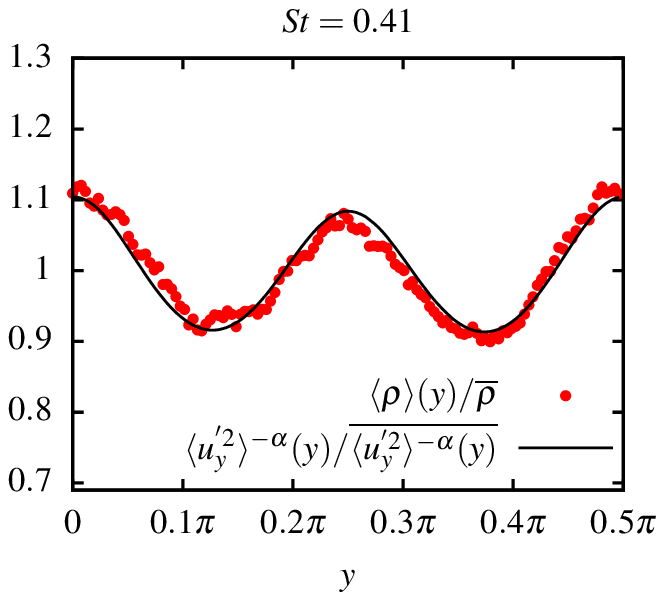}}

\vspace{-1.6em}\subfloat[]{
\includegraphics[scale=1]{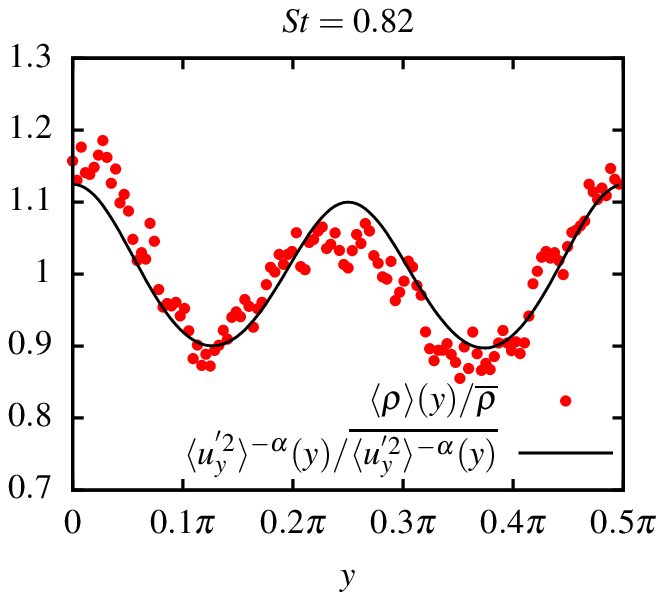}}
\caption{Comparison of normalized particle number density profiles $\lra{\rho}(y)/\overline{\rho}$ (red points) 
with the normalized profiles of transversal velocity fluctuations $\lra{u_y'^2}^{-\alpha}(y)/\overline{\lra{u_y'^2}^{-\alpha}(y)}$ 
(black solid line) for $Wi=23.9, Re=0.664$ and different values of $St$ (increasing from top to bottom). 
The values of the exponents obtained from a best fit are $\alpha= 0.229 \pm 0.029$ (a), $\alpha= 0.839 \pm 0.149$ (b), $\alpha= 1 \pm 0.309$ (c). 
The inset in (a) is a zoom around $\lra{\rho}(y)/\overline{\rho} = 1$.
All the shown profiles are further averaged over the wavelength defining the periodicity of the mean flow $\ell=LL_0=\pi/2$. 
}
\label{fig:rho-uysqr}
\end{figure}

The theoretical understanding of turbophoresis relies on statistical approaches. Models available in the literature 
are typically derived either from the Fokker-Planck equation obeyed by the probability density to find a particle at position 
$\bm{x}$ with velocity $\bm{v}$ at time $t$ (as in \cite{BFF14}), or on the application of a decomposition into mean and fluctuating 
components, in the spirit of Reynolds averaging, in fluid momentum and particle mass conservation equations (as in \cite{caporaloni1975}). 
Here we follow the second approach which, in spite of its more heuristic character, is perhaps more physically transparent; 
after a proper correspondence is made, both models provide the same results for what concerns the present discussion.  
We then write $f(\bm{x},t)=\lra{f}(y) + f'(\bm{x},t)$ for each quantity of interest $f(\bm{x},t)$, where the prime 
indicates the fluctuation. 
Defining as $\bm{J}=\rho \bm{v}$ the flux associated with the number density of particles, we have: 
\begin{equation}
\lra{J_y}(y) = \lra{\rho}(y) \lra{v_y}(y)+ \lra{\rho'v_y'}(y) 
\label{eq:Jy}
\end{equation}
for its component in the direction of inhomogeneity $y$. As is often done \cite{caporaloni1975,Guha1997} 
we adopt a gradient diffusion model for the second term on the right hand side of Eq. (\ref{eq:Jy}): 
\begin{equation}
\lra{\rho'v_y'}(y) = -D_p \frac{d}{dy}\lra{\rho}(y), 
\label{eq:graddiff}
\end{equation} 
where $D_p$ is the diffusion coefficient of the inertial particles. This is typically assumed to be close to that of fluid tracers 
(i.e. the eddy diffusion coefficient) $D_f$, which is completely justified only in the limit of vanishingly small 
Stokes number. Estimating $D_f$ dimensionally, one has:
\begin{equation}
D_p \approx D_f \approx \tau_c \lra{u_y'^2}(y),
\label{eq:Dp}
\end{equation} 
where $\tau_c$ is a correlation time associated with the fluid flow. 
We expect it to be proportional to $\tau_{\dot{\gamma}}$ (Eq.(\ref{eq:tau_gammadot})), so that $\tau_{\dot{\gamma}}/\tau_c=a$ 
with $a$ some constant of order 1. 
Still in the limit of $\tau_p \to 0$, using $\bm{v} \simeq \bm{u}-\tau_p(\partial_t \bm{u}+\bm{u}\cdot\bm{\nabla}\bm{u})$, 
the turbophoretic velocity in Eq. (\ref{eq:Jy}) can be expressed as
\begin{equation}
\lra{v_y}(y)  = -\tau_p \frac{d}{dy} \lra{u_y'^2}(y). 
\label{eq:turboph_v}
\end{equation}
Inserting Eq. (\ref{eq:graddiff}), with (\ref{eq:Dp}), and Eq. (\ref{eq:turboph_v}) into Eq. (\ref{eq:Jy}), 
for the fluxless steady state (i.e. $\lra{J_y}(y)=0$) we finally obtain:
\begin{equation} 
\lra{\rho}(y) \sim \lra{u_y'^2}^{-\alpha}(y),  
\label{eq:alpha_pdistr}
\end{equation} 
giving the relation between the inhomogeneities of the particle distribution and those of fluid velocity fluctuations. 
In this expression the exponent $\alpha = \tau_p/\tau_c=aSt$ controls the amplitude and shape of the spatial modulation 
of the particle density transversal profile. 
\begin{figure*}[]
\centering
\includegraphics[width=.8\textwidth]{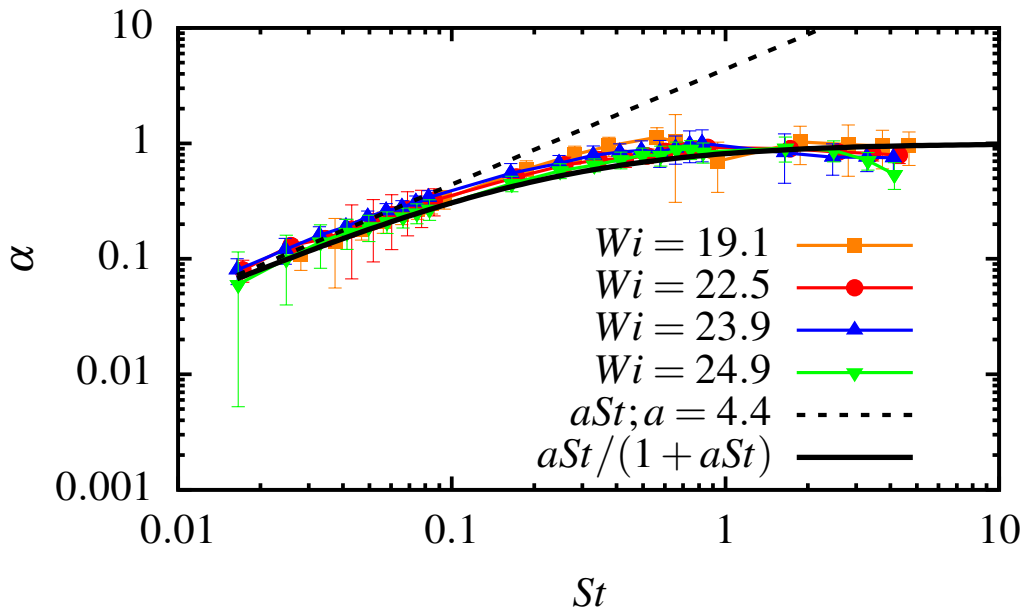}\hspace{2pc}%
\caption{Exponent $\alpha$ as a function of $St$ and for different Weissenberg numbers. The black dashed line represents the linear prediction 
in the limit of small $St$. The black solid line represents the modified non-linear prediction; $a=4.4$ is obtained by a best fit.
}
\label{fig:alpha}
\end{figure*}

The numerical results, shown in Fig. \ref{fig:rho-uysqr}, are in quite good agreement with the expectation of Eq. (\ref{eq:alpha_pdistr}), 
providing quantitative support to the claim that the large-scale inhomogeneities of the particle distribution are controlled by a 
turbophoretic mechanism. The small asymmetries observable in $\lra{\rho}(y)$ are due to the very slow convergence of particle statistics.  
As in Fig. \ref{fig:rho-trace-uysqr}, the results shown here are obtained by further averaging profiles over one forcing wavelength.  
The exponent $\alpha$, measured by a fitting procedure, is found to increase with the Stokes number and 
to approach $\alpha \simeq 1$ for $St \approx 1$ or larger (Fig. \ref{fig:alpha}). The growth of $\alpha$ with $St$ means that 
the amplitude of large-scale modulations of $\lra{\rho}(y)$, and hence the importance of turbophoresis, 
grows with increasing particle inertia.   
For the smallest values of $St$, $\alpha$ is found to linearly grow with $St$, with a value of the fitted 
proportionality constant $a=4.4$ (see the dashed black line in Fig. \ref{fig:alpha}). Hence, in this range 
of small particle inertia the numerical results are commensurate with the model prediction $\alpha \sim St$ 
valid in the limit of vanishingly small $St$.   
For larger $St$, the data are no longer described by this linear relation, with $\alpha$ tending to saturate to $1$. 
To account for this behavior we follow \cite{caporaloni1975} and \cite{Hinze1959}, where it was suggested that 
the shear-normal particle kinetic energy is different from the fluid one, being proportional to it through 
a $St$-dependent coefficient $\kappa$. 
The turbophoretic velocity in Eq. (\ref{eq:turboph_v}) should then be modified as follows:
\begin{equation}
\lra{v_y}(y) = -\kappa \tau_p\frac{d}{dy}\lra{u_y'^2}(y), 
\label{eq:turboph_v_kappa}
\end{equation}
where $\kappa = 1/(1+\tau_p/\tau_c)$ \cite{caporaloni1975}.
Reasoning as before, we obtain a fluxless steady solution like the one in Eq. (\ref{eq:alpha_pdistr}) but with 
\begin{equation}
\alpha = \frac{a St}{1 + a St}.
\label{eq:alpha_pdistr_nl}
\end{equation}
Remark that, from this, $\alpha \simeq a St$ for $St \simeq 0$ and $\alpha \to 1$ for very large $St$. This modified 
Stokes dependence captures quite well the behavior of the exponent $\alpha$ in a considerably broader range of $St$ 
extending to unity and beyond, as shown in Fig. \ref{fig:alpha} (solid black line, with $a=4.4$ as for the linear behavior). 
For even larger values of $St$ we were unable to obtain satisfactorily converged particle statistics. We note that these 
results weakly depend on $Wi$.  

\begin{figure*}[]
\centering
\includegraphics[width=.8\textwidth]{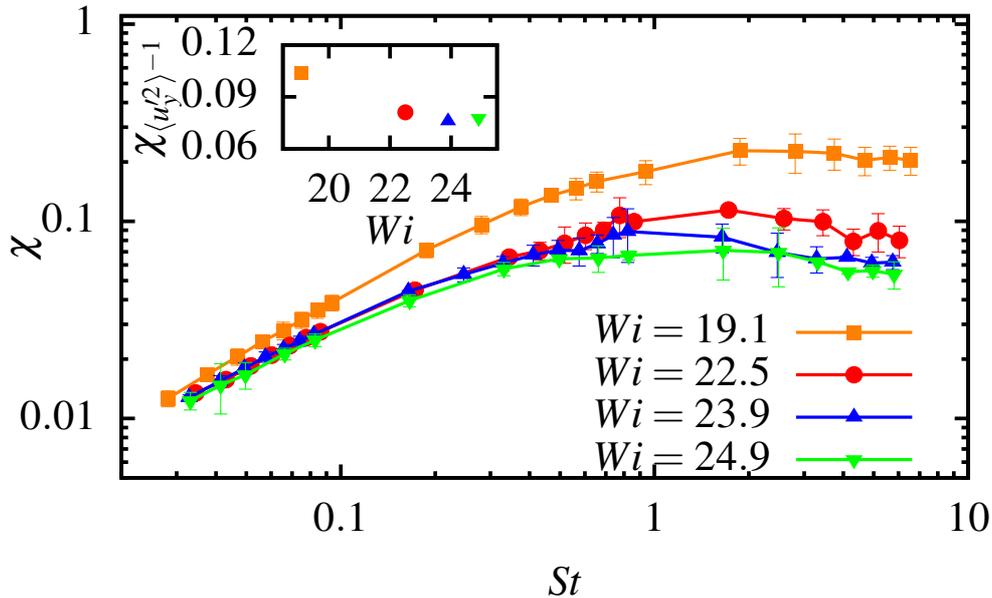}\hspace{2pc}%
\caption{Root-mean-square (rms) relative deviation $\chi$ of $\lra{\rho}$ from the mean uniform distribution $\overline{\rho}$ 
as a function of $St$ and for different Weissenberg numbers. 
Here temporal averages are performed over 80 independent realizations corresponding to different
instants of time separated by an interval larger than the typical flow time scale.
The inset shows the parameter $\chi_{\lra{u_y'^2}^{-1}}$ computed 
from the rms relative deviation of $\lra{u_y'^2}^{-1}$ from the mean uniform value $\overline{\lra{u_y'^2}^{-1}}$, 
i.e. $\displaystyle \sigma_{\lra{u_y'^2}^{-1}}/\overline{\lra{u_y'^2}^{-1}}$, 
as a function of $Wi$.}
\label{fig:chi} 
\end{figure*}

To quantitatively assess the overall effect of turbophoresis, as in \cite{DCMB16}, we measure the rms relative deviation 
of the mean particle density profile $\lra{\rho}(y)$ from the uniform distribution $\overline{\rho}=1/L_0$, defined as: 
\begin{equation}
\chi \equiv \frac{\sigma_{\lra{\rho}(y)}}{\overline{\rho}} = 
\left[ \frac{1}{L_0} \int_{0}^{L_0} \left(1-\frac{\lra{\rho}(y)}{\overline{\rho}}\right)^2dy \right]^{\frac{1}{2}}, 
\label{eq:chi_def}
\end{equation}
where $\sigma_{\lra{\rho}(y)}$ is the standard deviation of $\lra{\rho}(y)$. 
The global parameter $\chi$ as a function of Stokes for different $Wi$ numbers is presented in Fig. \ref{fig:chi}. 
Consistently with the behavior of $\alpha$, we find that $\chi$ grows with $St$ and eventually reaches an approximately 
constant value for $St \geq 1$. 
In the limit of $St \to \infty$, we would expect $\chi$ to be a decreasing function of $St$, due to the fact that,  
practically, very heavy particles should not interact with the flow field. However, this point could not be verified 
within this study, due to finite statistics associated with the difficulty to attain large enough Stokes numbers. 
The approximately constant behavior of $\chi$ for $St \geq 1$ appears nevertheless reasonable from its definition, considering 
that in the same range of Stokes numbers the exponent $\alpha$ characterizing $\lra{\rho}(y)$ is at a plateau value. 

Finally, we observe that $\chi$ displays some dependency on $Wi$, which becomes more evident as $St$ increases. 
Indeed, as seen from Fig. \ref{fig:chi}, $\chi$ decreases with increasing $Wi$, suggesting that 
the large-scale accumulation of particles (quantified by $\chi$) decreases with increasing polymer elasticity. 
A possible explanation of this trend can be the slightly flatter shape of $\lra{u_y'^2}(y)$ for increasing $Wi$ (not shown), 
corresponding to more homogeneous fluid velocity fluctuations, around its mean value (that, instead, grows with increasing 
$Wi$ since it represents the average intensity of transversal velocity fluctuations). 
If we focus on the region $St=O(1)$ where the effect of varying $Wi$ is most important, and 
we substitute $\lra{\rho}(y)$ with $\lra{u_y'^2}^{-1}(y)$ (notice that $\alpha \simeq 1$ for $St=O(1)$) in the expression 
of $\chi$, we then should have a decrease of its plateau value with $Wi$. 
As shown in the inset of Fig. \ref{fig:chi}, the computation of $\chi_{\lra{u_y'^2}^{-1}}$, i.e. the one based on $\lra{u_y'^2}^{-1}(y)$ 
confirms this expectation. 

\section{Conclusions}\label{sec:4}
The small and large scale inhomogeneities of the distribution of heavy inertial particles passively transported 
by a 2D elastic turbulence flow with (Kolmogorov) sinusoidal mean shear \cite{BCMPV05,BBBCM08,BB10} have been investigated 
by means of direct numerical simulations. 

A strong correlation bewteen the particle distribution and the polymer (square) elongation field was detected, with large 
particle concentrations occurring along thin highly elastic filamentary structures. 
Since the interaction between polymers and particles is not direct in the adopted model dynamics, but rather mediated 
by the fluid flow, it has been possible to interpret such a phenomenon in terms of the preferential concentration 
of particles outside vortices, in strain dominated regions where, in turn, polymers are efficiently stretched. 
The statistical features of small-scale clustering were further addressed measuring the correlation dimension of the fractal sets 
on which particles accumulate, i.e. the scaling exponent of the probability density to find particle pairs at small distances. 
The analysis revealed particularly effective clustering for Stokes numbers of order unity, for which $D_2$ decreases to approximately 
$1$, pointing to the aggregation of particles on almost one-dimensional structures. 
The considered statistics display only rather weak dependence on the Weissenberg number, in the range of parameters explored. 

At large scales, a turbophoretic mechanism associated with the gradients of eddy diffusivity was found to be responsible 
of segregation, as in Newtonian fluids at high Reynolds number \cite{picano2009spatial,Sardina-2012,DCMB16}. 
Indeed, the particle spatial distribution 
is strongly linked to the structure of the mean and fluctuating components of the fluid velocity, with maxima in correspondence 
to the minima of the shear-normal (elastic) turbulence intensity. A detailed analysis allowed us to measure the exponent 
characterizing the relation between the mean particle density profile and turbulence intensity in the direction transversal 
to the mean flow. 
Differently from the case of the 3D Newtonian turbulent Kolmogorov flow, this exponent was found to depend on particle inertia, 
i.e. on the Stokes number. 
Such a dependence resulted to be non-linear in $St$ and could be explained by adapting previous 
theoretical approaches \cite{caporaloni1975,BFF14} to construct a simple model by means of a Reynolds averaging procedure.   
A similar non-linear dependence is also reflected in the overall intensity of the turbophoresis phenomenon, quantified by the global 
parameter $\chi$ accounting for the rms deviation of the particle distribution, relative to the uniform one. 
This quantity shows some negative dependence on the Weissenberg number, suggesting a reduction of segregation 
for larger values of $Wi$,  a feature that is likely related to the progressively (with growing $Wi$) less inhomogeneous character 
of transversal fluid velocity fluctuations.  

These results were obtained adopting the constant viscosity Oldroyd-B model of viscoelasticity. 
While it has been argued that the main features of elastic turbulence are quite independent of the rheological model 
details \cite{FL03,BBBCM08,PGVG17,GP17}, the effect of the latter on particle dynamics might not be unimportant. Rheological 
models accounting for shear-dependent viscosity effects (such as FENE models) could bring in additional dynamical couplings 
between the flow and the particles \cite{NSPL13}. One could indeed expect that, e.g., in a shear-thinning fluid the varying 
effective viscosity would reduce the drag force experienced by the particles in the regions of the flow where polymers are 
maximally stretched, and this might in turn affect the particle unmixing properties. 
It is a subject that deserves future investigations in order to assess to what extent the phenomenology described in this paper would apply.

\section*{Acknowledgments}
The research leading to these results has received funding from European COST Action MP1305 ``Flowing matter''. 



\end{document}